\documentclass[prl,twocolumn,preprintnumbers,amsmath,amssymb]{revtex4}

\pdfoutput=1

\usepackage{epsfig,amsfonts}

\def\be{\begin{equation}}
\def\ee{\end{equation}}

\def\bea{\begin{eqnarray}}
\def\eea{\end{eqnarray}}

\def\e{\epsilon}

\def\a{\alpha}
\def\b{\beta}

\def\ben{\begin{enumerate}}
\def\een{\end{enumerate}}

\newcommand{\bra}[1]{\left< #1 \right|}
\newcommand{\ket}[1]{\left| #1 \right>}
\newcommand{\vm}[1]{\left< #1 \right>}

\begin{document}

\title{%Exact quantum quench dynamics of the fermionic pairing model}
Quantum quenches from integrability: \\
the fermionic pairing model}
\author{Alexandre Faribault${}^{1,2}$, Pasquale Calabrese${}^{3}$ and  
Jean-S\'ebastien Caux${}^{2}$}
\affiliation{$^1$
Physics Department, Arnold Sommerfeld Center for Theoretical Physics,
and Center for NanoScience, \\
Ludwig-Maximilians-Universit\"at, Theresienstrasse 37, 80333 Munich, Germany}
\affiliation{
$^2$Institute for Theoretical Physics, Universiteit van  
Amsterdam,
1018 XE Amsterdam, The Netherlands}
\affiliation{$^{3}$Dipartimento di Fisica dell'Universit\`a di Pisa and  
INFN,
56127 Pisa, Italy}

\date{\today}

\begin{abstract}
Understanding the non-equilibrium dynamics of extended quantum systems
after the trigger of a sudden, global perturbation (quench) represents a 
daunting challenge, especially in the presence of interactions.  The main
difficulties stem from both the vanishing time scale of the quench event,
which can thus create arbitrarily high energy modes, and its non-local
nature, which curtails the utility of local excitation bases.  We here
show that nonperturbative methods based on integrability can prove
sufficiently powerful to completely characterize quantum quenches:  
we illustrate this using
a model of fermions with pairing interactions (Richardson's model).
The effects of simple (and multiple) quenches on the dynamics 
of various important observables are discussed.  Many of the features we find 
are expected to be universal to all kinds of quench situations in atomic
physics and condensed matter.

\end{abstract}

\maketitle

The experimental realization \cite{exp} of cold atomic systems with a high 
degree of tunability of Hamiltonian parameters, and the ability to evolve in 
time with negligible dissipation,
has reignited the study of many-body quantum systems away from equilibrium.
How Gibbs or any other relaxed states can ultimately result from unitary 
dynamics is a question that has received a lot of attention 
recently \cite{th-mix,cc-06,gdlp-07,IS,s08}, but which still lacks a 
general understanding. %, at least for strongly-correlated systems.

Suppose an extended quantum system is prepared in one eigenstate
$|\psi_{g_0}^\mu\rangle$ of some Hamiltonian $H_{g_0}$, where $g_0$ is a 
tunable, global parameter (interaction strength, external field, ...).  
At a given time, say $t=0$, this parameter is suddenly changed to a different 
value $g$, and the system thus starts evolving {\it unitarily}
according to the dynamics governed by a different Hamiltonian $H_g$.
This is what is referred to as a quantum quench. 
The resulting time evolution is simply given by the solution of the 
Schr\"{o}dinger equation $|\psi(t)\rangle=e^{-i H_g t} |\psi_{g_0}^\mu\rangle$.
Since $|\psi_{g_0}^\mu \rangle$ is {\it not} an eigenstate of $H_g$, this
can be extremely difficult to quantify.
The most straightforward way to tackle the problem is therefore to write
the initial state $|\psi_{g_0}^\mu\rangle$ as a sum over the complete set of 
eigenstates $|\psi_{g}^\nu\rangle$ (having energy $\omega_g^{\nu}$) of 
Hamiltonian $H_g$, leading to the time-dependent post-quench state
\be
|\psi(t)\rangle=\sum_{\nu} e^{-i\omega_g^\nu t}
\langle\psi_{g}^\nu|\psi_{g_0}^\mu\rangle |\psi_g^\nu\rangle\,.
\label{Psi_of_t}
\ee 
The complexity of the problem  is encoded first in the distribution of 
energies $\omega^{\nu}_g$, but most importantly in the matrix of overlaps
of eigenstates pertaining to the two different Hamiltonians,
\be
Q_{g g_0}^{\nu\mu}\equiv \vm{\psi^\nu_{g}|\psi^\mu_{g_0}}\,,
\label{QuenchMatrix}
\ee
which we call the {\it quench matrix}. This matrix is of dimension
equal to that of the Hilbert space
\footnote{Formally, the pre- and post-quench Hamiltonians do not have to
share the same Hilbert space (a quench could for example be defined which would
kill off or introduce new degrees of freedom).  
The quench matrix thus technically has dimensions
$\mbox{dim}({\cal H}_g) \times \mbox{dim}({\cal H}_{g_0})$, and is well-defined
provided we adopt a proper measure for the scalar product.},
but in practice we mainly need a single column expressing 
the initial eigenstate of $H_{g_0}$ in terms of the eigenstates of $H_g$.
However, even in the very few cases when all eigenstates of 
a many-body Hamiltonian can be classified and written down, the calculation of 
the quench matrix coefficients is a severe challenge, whose 
computational complexity generally grows factorially with system size. 
%and/or number of excitations.
Shortcuts can be found for systems having a representation in terms of 
free particles (like the 1D Ising chains \cite{IS,s08}) where Wick's 
theorem suffices to calculate all the overlaps, but for truly interacting 
systems this remains a very ambitious programme.  
Most of the theoretical work on quantum quenches has up to now concentrated 
on the calculation of correlation functions in 
specific regimes \cite{th-mix,cc-06},
with little reference to the post-quench state of the system.

Besides describing the state resulting from a quench of state $\mu$, it is 
also important to be able to characterize the time dependence of physical 
observables $O$ after the quench.  Formally, we can write
\bea
\vm{O(t)} &\equiv& 
\langle \Psi (t) | O | \Psi(t) \rangle \nonumber \\
&=& \sum_{\nu,\xi} e^{i (\omega_g^\nu-\omega_g^\xi) t}
Q_{g_0 g}^{\mu \nu} Q_{g g_0}^{\xi \mu} \bra{\psi^\nu_g}O\ket{\psi^\xi_g}\,,
\label{quenchO}
\eea
where calculating the matrix elements $\bra{\psi^\nu_g}O\ket{\psi^\xi_g}$ 
represents an additional hurdle for interesting observables in 
nontrivially interacting models.  
Even if we are able to obtain these matrix elements, the leftover double sum 
over the full Hilbert space is enormous, and one can wonder whether this way 
of proceeding is of any practical use.
New nonperturbative methods are clearly needed to obtain a proper description
of the physics involved.

The purpose of the present paper is to introduce a new line of attack on 
quantum quench problems, sufficiently powerful to yield not only the quench 
matrix of specific interacting problems (and thus the ensuing nonequilibrium 
state), but also able to provide matrix elements of physical observables, 
and thus their time dependence after the quench.  
This approach is based on the exact solvability of certain many-body quantum 
problems known as integrable or Bethe Ansatz \cite{BetheZP71} 
solvable theories.  
Integrability came into prominence as a means of obtaining exact results for 
the equilibrium thermodynamics of one-dimensional systems (see 
\cite{KorepinBOOK,TakahashiBOOK} and references therein).  
%These matrix elements are of great utility, mainly in the computation
%of equilibrium correlation functions.
More recently, a description of correlation functions at
equilibrium has been achieved by exploiting results from the Algebraic Bethe
Ansatz (ABA), which provides economical expressions for matrix elements of 
local operators in the basis of exact Bethe eigenstates.  
The existence of these expressions stems from two results:
Slavnov's formula \cite{s-89} for the overlap of a Bethe state with a generic
state, and the solution of the so-called quantum 
inverse problem \cite{KitanineNPB554},
{\it i.e.} the mapping of physical operators to ABA operators.
These matrix elements are of great utility in the computation
of equilibrium correlation functions.
%Summing these matrix elements over intermediate states then
%grants access to the dynamical correlation functions.  
One very important feature %of such an approach 
is that Bethe states typically 
offer a very optimized basis in which only a very small minority of
eigenstates carry substantial correlation weight, allowing the summation over
intermediate states to be drastically truncated without significantly 
affecting the results. %saturation of the sum rules.
This novel approach has been successfully applied to equilibrium correlations
of quantum spin chains \cite{cm-05} and atomic Bose gases \cite{cc-06b}.
We here further extend the reach of integrability into the domain of 
non-equilibrium quench dynamics.

\section*{THE MODEL AND ITS SOLUTION}

We consider a model of spin $1/2$ fermions in a shell of 
%single-particle 
energy levels $\e_\a$ with a Cooper pairing-like interaction
\be
H = \sum_\a\sum_{\sigma} \frac{\e_\a}{2} c^{\dagger}_{\a \sigma}
c_{\a \sigma} - g \sum_{\a, \beta} c^{\dagger}_{\a+}c^{\dagger}_{\a-}
c_{\beta-} c_{\beta+}\,,
\label{eq:H_fermions}
\ee
which was introduced by Richardson \cite{rs-62} in the context of nuclear 
physics, and has found applications in the physics of ultrasmall metallic 
grains \cite{rev}. It reduces to conventional Bardeen-Cooper-Schrieffer (BCS) 
theory \cite{bcs-57} in the thermodynamic (TD) limit.  
The model has a pseudo-spin representation with 
$S^-_\a= c_{\a\downarrow}c_{\a\uparrow}$ (see Methods) with $N$ spins.
Central to our approach is the fact that this Hamiltonian can be diagonalized 
using the Bethe Ansatz \cite{rev,zlmg-02}. 
The Hilbert space separates into sectors 
%of different magnetizations, or equivalently 
of fixed number of down spins $N_r$.  
The eigenstates of the model are given by Bethe wavefunctions, 
each individually characterized by a set of $N_r$ rapidities $\{ w_j \}$ 
obeying a set of algebraic equations known as the Richardson equations
\be
\frac1g=\sum_{\a=1}^N\frac1{w_j - \e_\a}-\sum_{k\neq j}^{N_r}  
\frac2{w_j-w_k}\,\quad
j=1,\dots, N_r\,,
\label{RICHEQ}
\ee
and are obtained by repeated action of an operator $B(w_j)$ on the fully 
polarized reference state $|0\rangle$:
\be
|\{w_j\}\rangle=\prod_j B(w_j)|0\rangle=
\prod_{k=1}^{N_r} \sum_{\a=1}^N \frac{S_\a^-}{w_k - \e_\a} |0\rangle.
\label{RICHWF}
\ee
The $\binom{N}{N_r}$ different solutions to (\ref{RICHEQ}) then allow to 
construct a full set of orthogonal eigenstates, providing us with a proper 
basis of the Hilbert space.

In Bethe Ansatz solvable models, Slavnov's formula \cite{s-89}, 
gives the overlap of an eigenstate $|\{w\}\rangle $ with a general Bethe 
state $|\{v\}\rangle$ (see Methods).
The main difference with other models solvable by Algebraic Bethe Ansatz
(like the one-dimensional Bose gas or the XXZ chain) is that in the 
definition of the eigenstates (\ref{RICHWF}), the coupling constant $g$ enters 
only implicitly through the solutions of the Richardson equation for $w_j$. 
Consequently, for this model, Slavnov's formula is enough to calculate the 
overlaps between two generic states at any coupling.  
For other models, where $B(w_j)$ depends explicitely on $g$, 
it can only be used between states defined by the same operators $B(w_j)$ 
and therefore the same $g$,
and a more general expression for the overlaps remains to be found.

We can then exploit the accessibility to the quench matrix for the Richardson 
model to show how useful integrability can be when studying quenches. 
However, before entering in the details of the quantum dynamics, it is 
important to remember that in the TD limit $N_r,N\to\infty$ at fixed filling, 
the dynamics becomes classical \cite{bl-06,bcs} because of suppression of 
quantum fluctuations. In this limit, %the state of the system and 
the dynamics of the canonical order parameter can be obtained 
analytically by exploiting classical integrability \cite{bl-06,bcs}. 
The framework we propose in this letter works in the mesoscopic regime 
(finite $N$), allowing to study the effects of quantum fluctuations. 
With this tool at hand, we can characterize in an exact manner the 
crossover taking place between microscopic and macroscopic physics, 
a task impossible to achieve with thermodynamical approaches.

We will argue in the following that, similarly to what is observed for
equilibrium correlation functions, only a relatively small set of states 
contributes significantly to the decomposition of the initial state in 
the new eigenbasis. The natural approach is then to truncate in an
optimal way the Hilbert space, so that a faithful representation of 
the initial state is obtained. The induced truncation error is easily 
evaluated looking at how close   
$\sum_\mu \left|\langle\psi_{g_0}^0|\psi_g^\mu\rangle\right|^2$ 
is to the desired value of $1$.
Using the truncated Hilbert space we can then calculate any observable 
or correlation function by brute force summing the relevant contributions. 

Compared to other numerical truncation methods, this approach has the great
advantage that time enters only as a parameter.  
The explicit expression (\ref{Psi_of_t}) for the wavefunction means that at 
any time, expectation values can be computed without knowing the previous 
history of the system (apart from the initial state) and there is therefore no 
accumulation of errors as time passes. On the other hand, compared to
numerical exact diagonalization, matrix elements %for many integrable models 
can be expressed using Slavnov's formula as matrix determinants whose 
computational complexity is algebraic and not exponential in the system size 
and/or number of excitations, thereby allowing to reach large system sizes.

\section*{NUMERICAL RESULTS}

\begin{figure}[t]
\includegraphics[width=8.5cm]{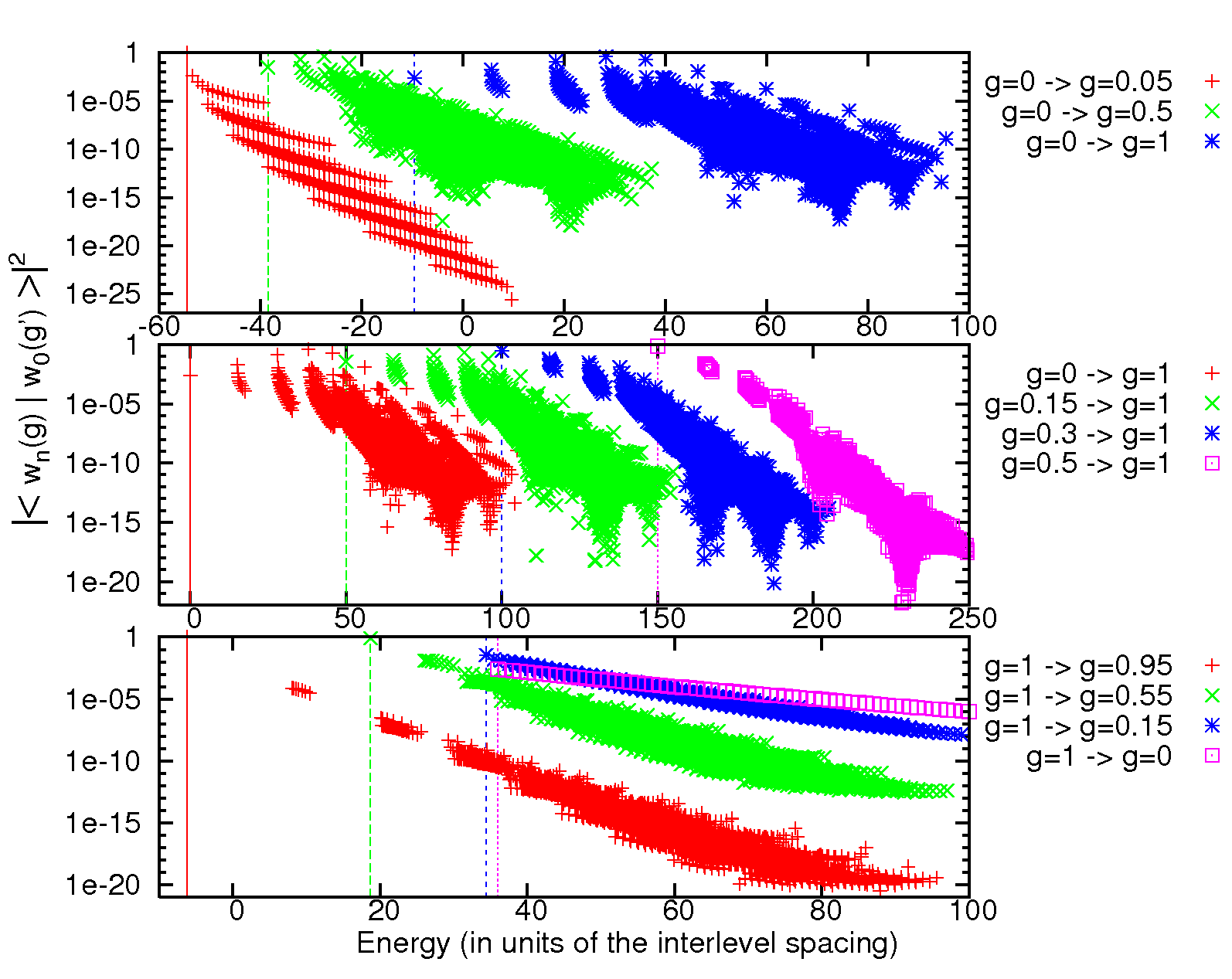} 
\caption{First column of the quench matrix (ground-state overlaps) for several
quenches. In all plots $N =2N_r= 16$ and the ground state energies (represented
by vertical lines) have been shifted for clarity.
Top: Decomposition of the $g=0$ ground-state with states at $g=0.05,0.5,0.95$. 
Center: Decomposition of several initial ground-state $g_0=0,0.15,0.3,0.5$
in terms of the states at $g=1$.
Bottom: Decomposition of the $g_0=1$ ground-state in terms of 
$g=0.95,0.55,0.15, 0$ states.
}
\label{quenchmat}
\end{figure}

\begin{figure}[t]
  \parbox{2.5cm}{
  \footnotesize
\bea
\bullet\bullet\bullet\bullet\bullet\bullet | \circ\circ\circ\circ\circ\circ
\nonumber\\
\bullet\bullet\bullet\bullet\bullet\circ | \bullet\circ\circ\circ\circ\circ
\nonumber\\
\bullet\bullet\bullet\bullet\circ\circ | \bullet\bullet\circ\circ\circ\circ
\nonumber\\
\bullet\bullet\bullet\circ\circ\circ | \bullet\bullet\bullet\circ\circ\circ
\nonumber\\
\bullet\bullet\circ\circ\circ\circ | \bullet\bullet\bullet\bullet\circ\circ
\nonumber\\
\bullet\circ\circ\circ\circ\circ | \bullet\bullet\bullet\bullet\bullet\circ
\nonumber\\
\circ\circ\circ\circ\circ\circ | \bullet\bullet\bullet\bullet\bullet\bullet
\nonumber
\eea
}
  \parbox{5.5cm}{
  \includegraphics[width=4.5cm]{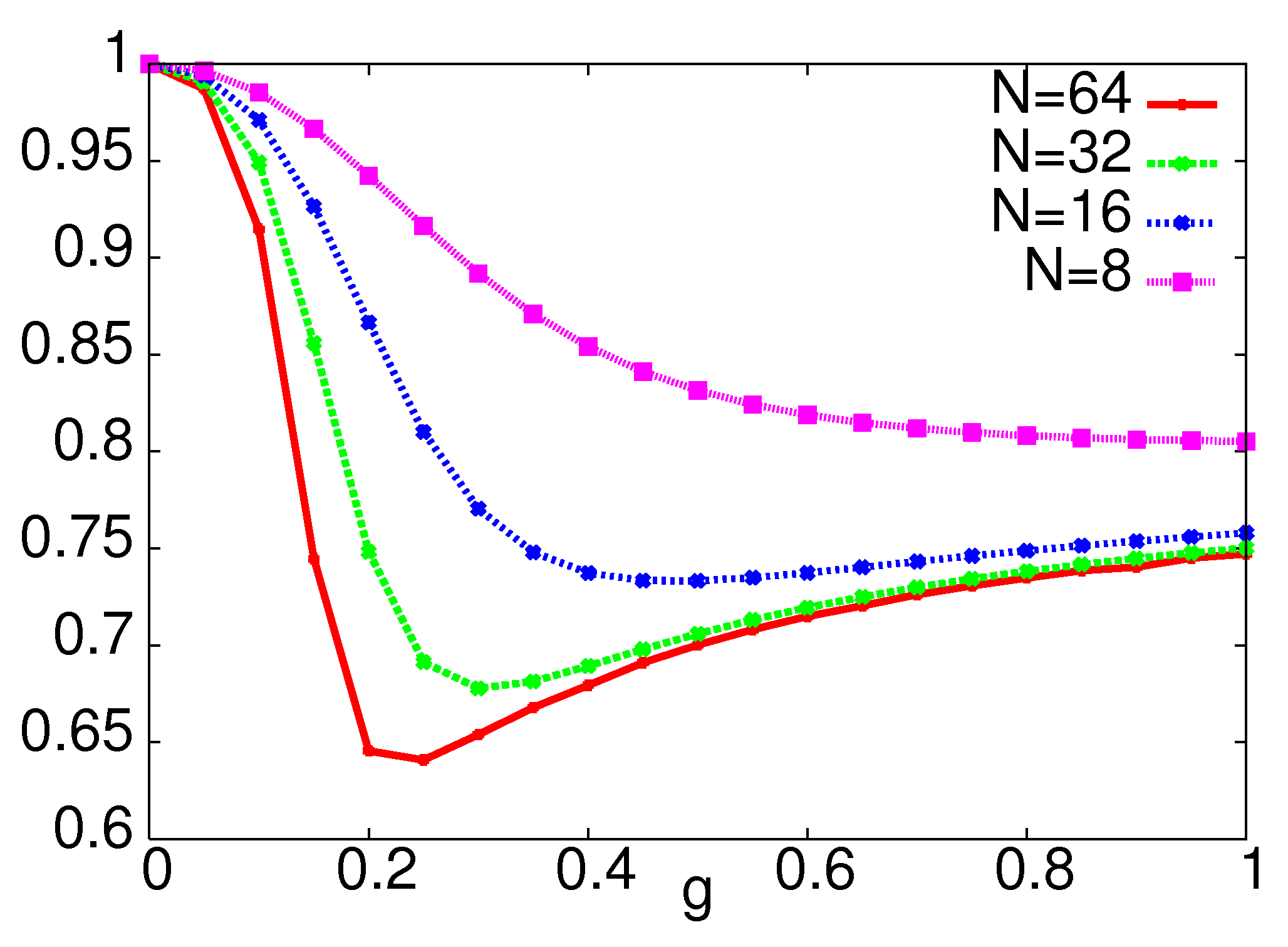}
}
\caption{Left: Pictorial representation of single block states 
obtained by promoting contiguous blocks of $N_p \le N_r$ rapidities 
from right below to right above the Fermi level.
Right: Total contribution of these states %$\in \mathcal{M}_{0,0}$ 
to the amplitude of the initial state at $g_0=0$, as a function of 
interaction.}
\label{M00}
\end{figure}

Let us now report explicit results starting with the quench matrix itself.
We concentrate on the case of equally spaced levels $\e_\a$ at half-filling, 
%i.e. $N_r=N/2$, 
but the method is clearly not limited to this case. 
Let us start from $N=2N_r=16$, when the Hilbert space has a dimension 
of ``only'' 12870. 
We can then numerically compute the rapidities for every single state. 
We report in Fig. \ref{quenchmat} the square of the overlaps 
for several quenches as obtained from Slavnov's formula (see Methods).
Starting from the non-interacting ($g=0$) ground state, the top inset 
shows the overlaps with all the states at three different finite couplings. 
This allows to understand some general features: %of the quench matrix:
having access to the complete quench matrix,
one realizes that only a few of the eigenstates at coupling $g$ have 
a large contribution to $|\psi_0\rangle$.
Therefore, getting a nearly exact description of the dynamics requires only 
a small subset of the states.
The final state having the largest overlap with the initial ground state is 
always one of the states 
%$\ket{N_p;g}$ 
that at $g=0$ is built 
%(and solved numerically for finite $g$) 
by flipping from up (down) to down (up) the $N_p$ spins 
right below (above) the Fermi level. 
We refer to these $N_r+1$ states as `single block states'. 
For the cases we studied here, quenching from a non-interacting initial state, 
these states always contribute more than $60\%$ of the total amplitude 
%of the states after the quench 
(see Fig. \ref{M00}). 
The total contribution of single block states is non-monotonic in $g$ showing 
(after decline at small interaction) a rise as $g$ gets sufficiently large.
%Notice the decreasing of their contribution when the system gets larger.

The `band-like' structure of the overlaps in Fig. \ref{quenchmat} makes it 
reasonable to assume that additional large contributions can be found for 
states built by slightly deforming the single block ones, e.g. by adding 
a single particle-hole excitation either above or below the Fermi level. 
The remaining most relevant states will then be those with two additional excitations.
%and so on. 
Following this assumption inductively, we add to the truncated Hilbert space 
multiple block states obtained by slightly deforming single block ones. 
This allows us to describe larger systems while 
retaining a tractable number of states. This procedure works extremely well, 
e.g. at $N=32$, for all the quenches from $g_0=0$ to $g\in (0,1]$, 
we were always able to find at minimum $97\%$ of the weight of the
initial state by using only 7000 states, i.e. only  $1/10^{5}$ of the full Hilbert space. Moreover, for a given final value of $g$, less than a 1000 of these states gives an actual important contribution.

In the center of Fig. \ref{quenchmat}, we report the overlaps obtained 
by quenching from different initial values of $g_0$ to the same final $g=1$. 
We see that the same band-like structure is present as when starting from 
a non-interacting state, leading to no qualitative change of the dynamics.
Vice-versa, the structure of the quench matrix for a reversed quench, i.e. 
from large to small $g$, reported in the bottom of  Fig. \ref{quenchmat}
is different: the weight of the state goes down exponentially with the energy 
of the states (almost straight lines in the figure), and for large quenches
the decay rate in energy is slow. An adequate representation of the initial state therefore requires that lots of states be taken into account. In this case, the optimal truncation procedure is still easily defined by simply keeping a sufficient number of low energy states.
 
%Although in this case, the optimal way to truncate the Hilbert space would be 
%to simply keep the lowest energy states, 
%the sheer number of important contributions rapidly becomes a limitation. 

\begin{figure}[t]
\includegraphics[width=8.5cm]{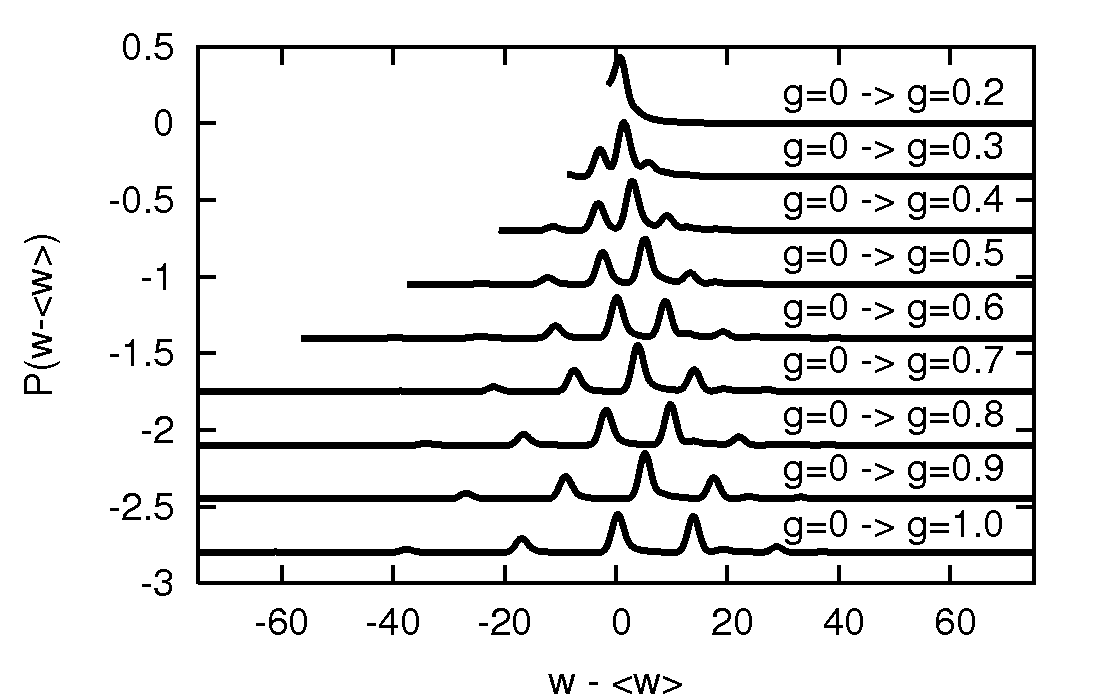} 
\includegraphics[width=8.5cm]{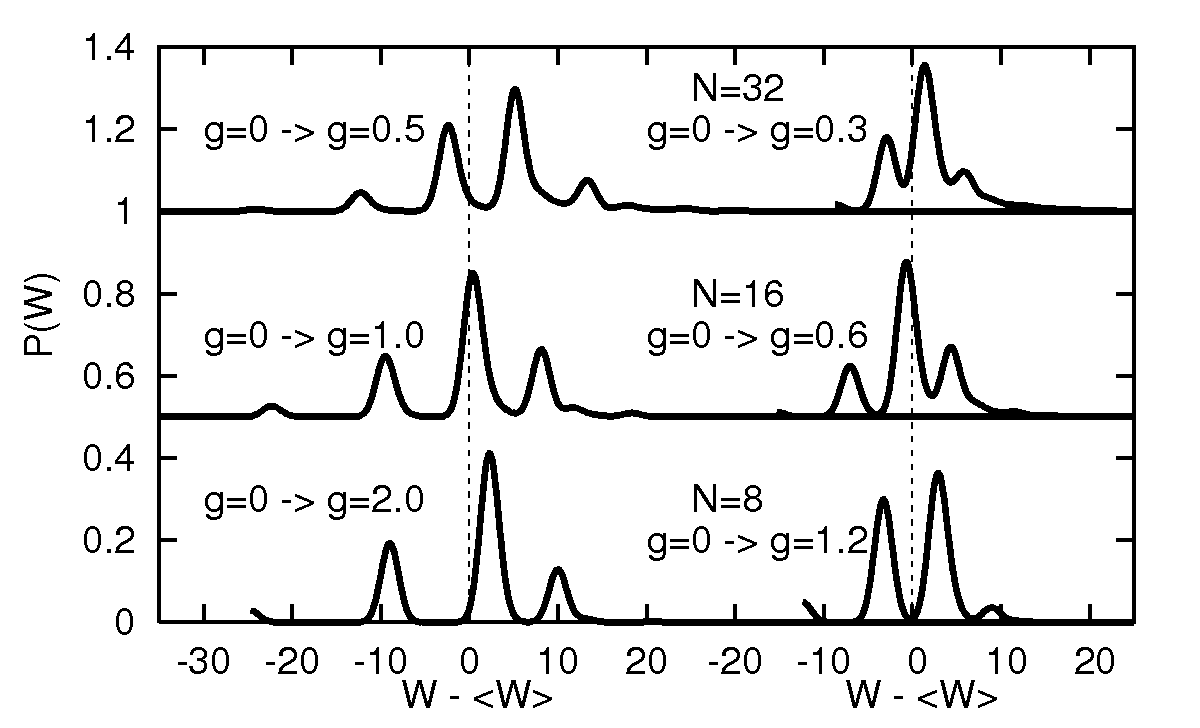} 
\caption{Distribution of work $P(W)$. %as function of $W-\langle W \rangle$.
Top: Different quenches at fixed $N=32$, showing the formation of multiple 
peaks for large quenches.
Bottom: At different $N$ keeping fixed $g N$.}
\label{PWfig}
\end{figure}

The first measurable quantity easy to derive from the knowledge of the quench 
matrix is the probability distribution of the `work' \cite{s08} 
\be
P(W)=\sum_\mu |\langle\psi_{g_0}^0|\psi_g^\mu\rangle|^2 
\delta(W-\omega_g^\mu +\omega_{g_0}^0)\,,
\ee
reported for some quenches from $g_0=0$ in Fig. \ref{PWfig}. 
These figures have been obtained by smoothing the $\delta$ function with a 
Gaussian of width proportional to the inter-level spacing. 
It is straightforward to derive
the average and the width of this distribution 
\bea
\langle W \rangle&=
& \langle \psi_{g_0}^0 |(H_g-H_{g_0}) |\psi_{g_0}^0\rangle
=(g_0-g) N_r \Psi_{OD}^{g_0}\,, \\
\langle W^2 \rangle&=& 
\langle \psi_{g_0}^0 |(H_g-H_{g_0})^2 |\psi_{g_0}^0\rangle=
(g_0-g)^2 N_r^2\Psi_2^{g_0}\,, \nonumber
\eea
where $\Psi_{OD}^{g_0}=
\langle\psi_{g_0}^0|(\sum^N_{\a,\b=1}S^+_\a S^-_\b)/N_r|\psi_{g_0}^0\rangle$ 
is the off-diagonal order-parameter in the initial 
state, %(which has been calculated for generic coupling \cite{fcc-08}), 
and $\Psi_2^{g_0}=\langle\psi_{g_0}^0|(\sum^N_{\a,\b=1}S^+_\a
S^-_\b/N_r)^2|\psi_{g_0}^0\rangle $ is a four point correlator. 
%that can be obtained from the form factors \cite{fcc-08}.
Higher cumulants are similarly obtained and only depend on the initial state
\footnote{In passing, we note that these relations also offer further sum 
rules connected to the conservation of energy.  In the truncated approach 
we used, these are well saturated.}.
%Note that 
%In the limit $N\to\infty$, the square factorizes i.e. 
%$\Psi_2^{g_0}= (\Psi_{OD}^{g_0})^2$, and the leading term in the 
%width of the distribution $\langle (\Delta W)^2\rangle$ vanishes.
%is proportional to $N$ goes to zero like $N^{-1}$ and the 
%As a consequence 
From TD relations \cite{s08}, it is generically expected that the probability 
of work per spin $w=W/N$ becomes a delta function. 
For finite $N$, $P(W)$ is non-trivial:
it shows a dominant peak close to $W=\langle W\rangle$, %which
%develops into the delta function with increasing $N$, 
but with %at finite $N$ shows 
a structure dictated by the presence of the state with the right quantum 
numbers at the given energy. 
In the top panel we report several quenches at $N=32$,
where the formation of subdominant peaks is explictly shown.
In the bottom panel we show $P(W)$ at different $N$ keeping $g N$ fixed. 
It is evident that despite the fact that the structure changes
drastically with $N$, the width of the distribution is constant,
indicating that, when written in terms of $W/N$, it becomes a delta function.

\section*{ORDER PARAMETER EVOLUTION}

\begin{figure}[t]
\includegraphics[width=8.5cm]{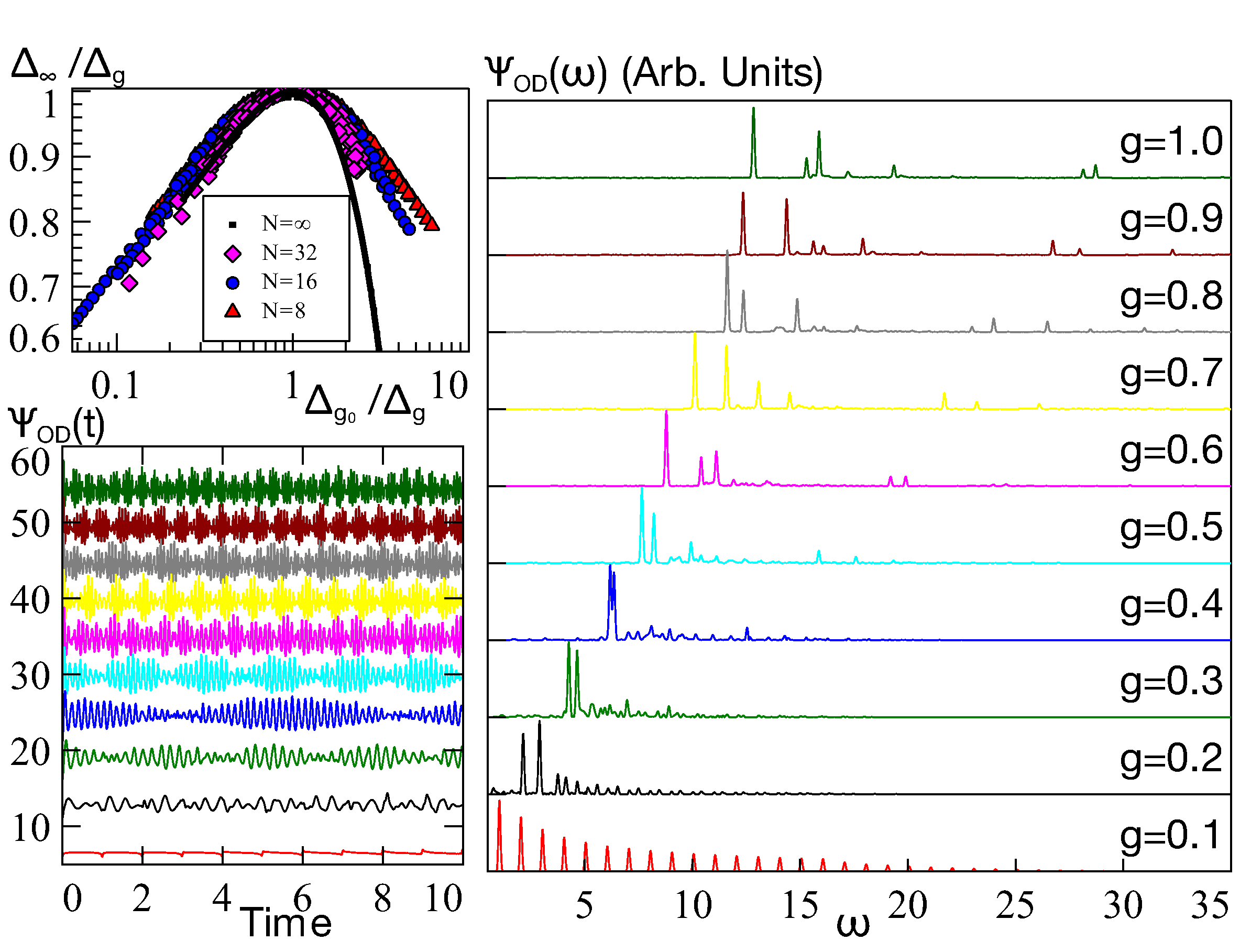}
\caption{Bottom left: Off-diagonal order parameter evolution for $N=32$. 
Right: Fourier transform,
the various plots are shifted on the vertical axis for clarity.
Top Left: Non-equilibrium finite-size ``phase diagram'' resulting from the 
time-averaged canonical gap obtained from the off-diagonal order parameter,
as explained in the text.}
\label{OP}
\end{figure}

We now present results for observables, concentrating on the %canonical 
off diagonal order parameter defined as  
\be
\Psi_{OD}(t)=
\langle\psi(t)|\frac1{N_r}\sum^N_{\a,\b=1}S^+_\a S^-_\b|\psi(t)\rangle\,.
\ee
In the equilibrium canonical ensemble $\Psi_{OD}$ for $N\to\infty$
is related to the BCS gap and so it is a natural quantity to understand
the superconducting tendency even out of equilibrium (on the same footing as 
the canonical order parameter used in \cite{bl-06}). 
According to Eq. (\ref{quenchO}) we can write the time evolution once 
the form factors for $\sum^N_{\a,\b=1}S^+_\a S^-_\b$ are known.
They have a representation in terms of a sum of $N_r$ determinants of 
$N_r\times N_r$ matrices depending on the rapidities 
%, which can be evaluated numerically 
(see Methods). 
%However, the numerical calculation can be time consuming
%because of the large number of determinants that have to be computed, 
%even in a heavily truncated Hilbert space. 
In the bottom left part of Fig. \ref{OP} we present the resulting 
real-time evolution of $\Psi_{OD}(t)$ starting from $g_0=0$ and evolving 
with several different $g$ for $N=32$. 
The information contained here is better extracted from the 
Fourier transforms reported on the right of Fig. \ref{OP}.

For small value of $g$, the various frequencies entering are very close to 
integers, as a result of the almost perfect equispacing of the levels. 
This regime could simply be described by perturbation theory and doesn't 
%necessarily 
show any striking features different from free fermions. 
With increasing $g$, the spectrum becomes very complicated since a  
large number of incommensurate frequencies contribute to the order parameter
evolution. 
This is the realm of quantum fluctuations which 
makes the evolution highly irregular.
Still increasing $g$, some regularity appears again.
This can be understood in terms of results in the TD
limit \cite{bl-06}. In fact, for $N\to\infty$, quenching from weak interaction 
to a much larger one, leads to an order parameter which shows persistent 
non-harmonic periodic evolution, i.e. a Fourier transform with equispaced 
peaks. Within the canonical description presented here, this feature will 
be reproduced when quenching to a large final value of $g$ since, as was shown 
in \cite{largeg}, excited states in this regime form equispaced bands at energy
$E\approx \Delta N_G+O(N^0)$  ($N_G$ being the number of Gaudinos, 
i.e. the relevant excitations in this regime \cite{largeg}). 
%Also any other observable would show this periodicity in time.
As can be readily seen looking at the energy distribution of the points in 
Fig. \ref{quenchmat}, for finite large couplings, the low energy bands 
are already clearly formed, progressively collapsing
into a single energy $E \approx \Delta N_G$. 
The slight remaining width of these bands would only result in additional low 
frequency corrections to the mean-field BCS result.

The static correlation functions studied in \cite{fcc-08,ao-02} depend mainly 
on low energy properties, and BCS-like behavior was always found 
when $g\agt g^*=(2\ln N)^{-1}$, i.e. the criterion for the presence of 
superconductivity.  
In the problem at hand though, the quench matrix clearly shows that quantum 
quenches lead to an important occupation of the higher energy bands. 
These clearly differ from the BCS spectrum even for values of $g$ much larger 
than $g^*$ (that for $N=32$ is only $0.144\dots$).  
Quenches, since they probe high energy properties of the system, open up 
%a large range of parameters for which 
the possibility of probing 
interaction effects not captured by the mean-field treatment. 
%can dominate the dynamics.
Non mean-field features are manifest in non-equispaced dominant peaks in 
the frequency dependence of the order parameter. 
These accessible experimental quantities could thus be used as a spectroscopic
tool (as proposed for other models in Ref. \cite{gdlp-07}) to study quantum
fluctuations. 

We also considered the evolution of the canonical order parameter defined as
$\Psi(t) =  \sum_{\a=1}^N \sqrt{\frac{1}{4} - \vm{S^z_\a(t)}^2}$, using the 
knowledge of the form factors for $S_\a^z$ (see Methods).
It displays the same qualitative features as $\Psi_{OD}(t)$ 
and consequently will not be discussed here.

Let us conclude this section with a discussion of the long time asymptotic. 
It is difficult to extract any information about it
from the highly irregular and oscillatory behavior reported 
in Fig. \ref{OP}. Furthermore in finite system, (approximate) quantum 
recurrence will always spoil any signature of an eventual asymptotic state. 
However, if in the TD limit the asymptotic value of an observable
exists, it must be equal to its time average, that is straightforwardly
obtained with the tools at hand in finite systems.
In fact, in Eq. (\ref{quenchO}) all the terms with $\nu\neq\xi$ average to 
zero and so $\overline{O}= 
\sum_\nu |Q^{0\nu}_{g_0 g}|^2 \langle\psi^\nu_g|O|\psi^\nu_g\rangle$,
(the overline stands for the time average). 
$\overline\Psi_{OD}$ is obtained with little effort using the
Hellmann-Feynman theorem 
$\langle\psi^\nu_g|\sum^N_{\a,\b=1}S^+_\a S^-_\b|\psi^\nu_g\rangle=
-\frac{\partial \omega^\nu_g}{\partial g}$, thus without involving determinants. 
The non-equilibrium `phase diagram' for $N\to\infty$ \cite{bl-06} 
shows that the final asymptotic value of the canonical gap $\Delta_\infty$ 
defines a universal curve when expressing $\Delta_\infty/\Delta_g$ 
vs $\Delta_{g_0}/\Delta_g$, where $\Delta_g$ is the equilibrium value 
at coupling $g$. 
In our normalization, $\Delta_g= g\Psi/N_r= g\sqrt{\Psi_{OD}/N_r -1}/N_r$ 
\cite{fcc-08} (the $-1$ cancels the first correction for large $N_r$). 
We take the last equation also off-equilibrium for the definition of the 
time averaged canonical gap in finite size. 
The resulting `finite-size phase diagram' is reported in 
Fig. \ref{OP} (top left),
where most of quenches from $g_0$ to $g$ both in $[0,1]$ are shown, 
%at step of $0.1$ 
for $N=8,16,32$ at half filling (we excluded the points with an equilibrium 
$\Psi_{OD}$ much different from BCS prediction, that are not expected to
approach the asymptotic result).
It is evident how increasing $N$ the curves tend to the Barankov-Levitov
result shown as a full line \footnote{Notice that for 
$\Delta_{g_0}/\Delta_g< e^{-\pi/2}$ in Ref. \cite{bl-06} a branch point is
predicted resulting in an oscillatory behavior. We limit
ourself to consider the average}.

\begin{figure}[t]
%\parbox{4.1cm}{\includegraphics[width=4cm]{Sequential_quench_NP.eps}}
\parbox{4.1cm}{\includegraphics[width=4cm]{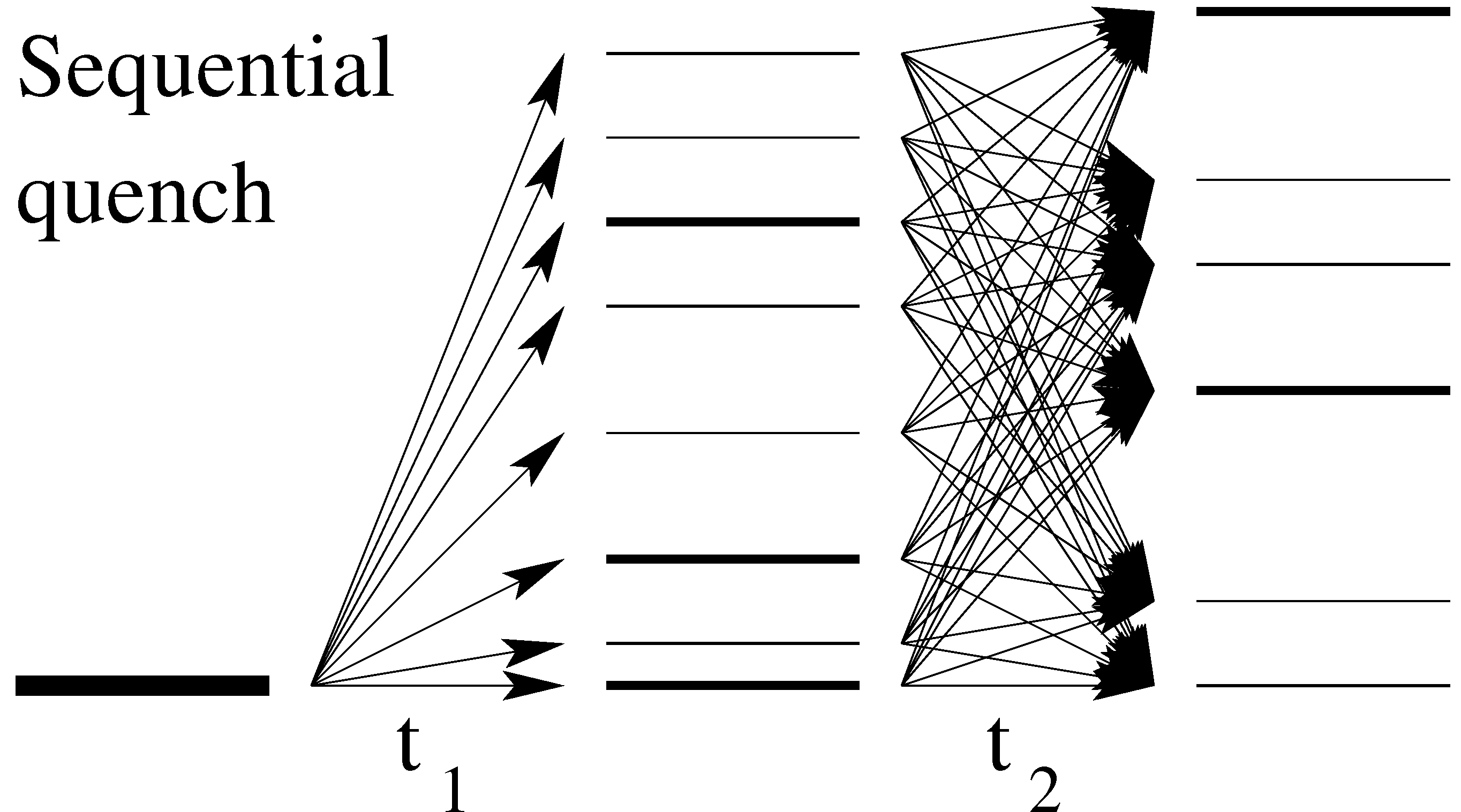}}
%\parbox{4.1cm}{\includegraphics[width=4cm]{Sequential_quench_localized_NP.eps}}
\parbox{4.1cm}{\includegraphics[width=4cm]{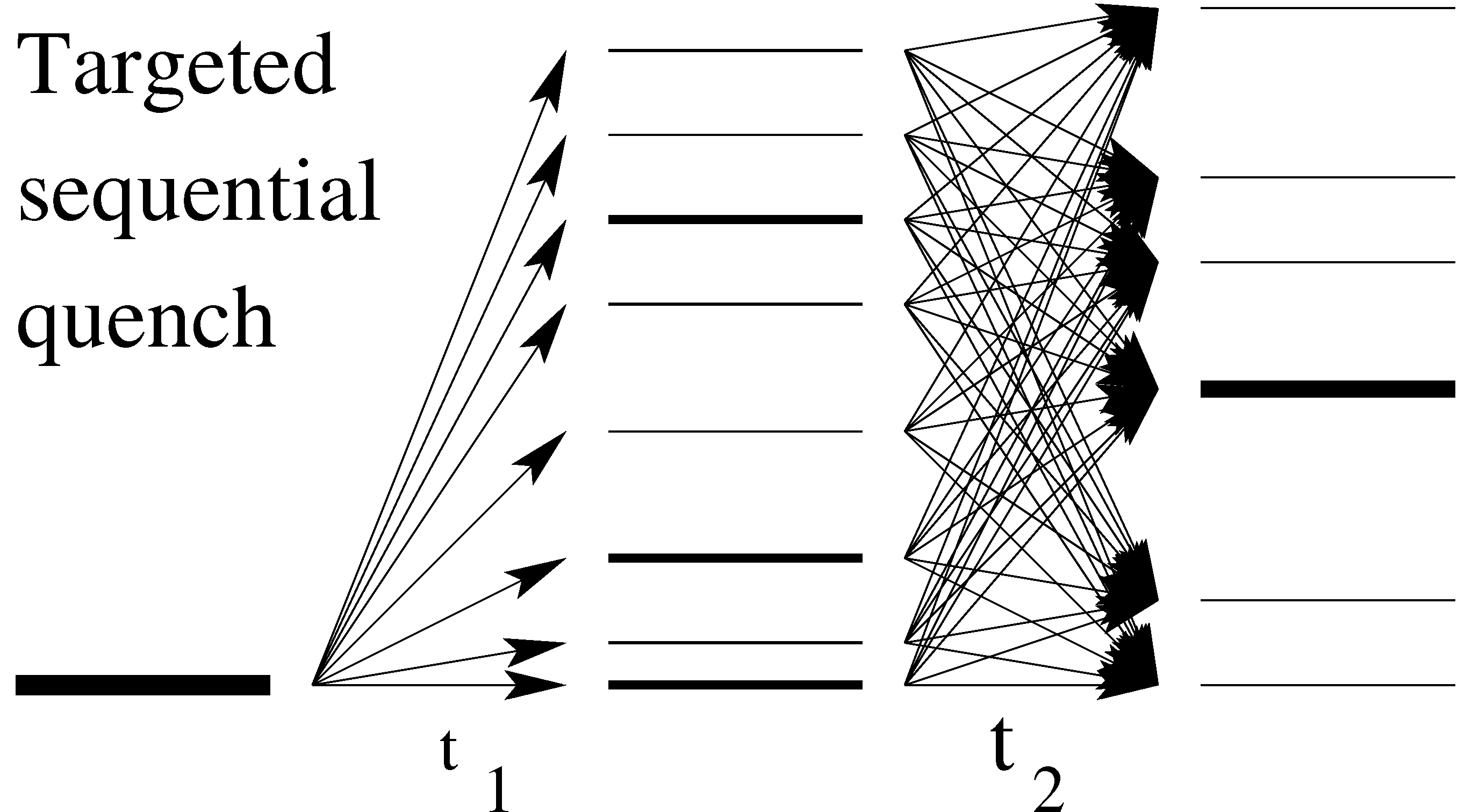}}
\caption{Left: A typical quench $g_0\to g$ populates all states according
  to the quench matrix. A general dequench $g\to g_0$ redistributes
the weight among all states. 
Right: A targeted dequench after a chosen time can populate a targeted state.}
\label{DQidea}
\end{figure}

\section*{THE DOUBLE QUENCH}

We move now to address the very interesting dynamics that appear when we 
consider sequences of multiple quenches, $g_i \to g_{i+1}$ at times $t_i$.
For brevity, we concentrate here on the %fundamental 
problem of the double quench, or quench-dequench sequence, defined by 
$g = g_0$ for $t < 0$, $g = g_1$ for $0 \leq t < t_q$ and $g = g_0$ for 
$t \geq t_q$.  
Starting from a specific eigenstate of $H_{g_0}$, 
the quench at $t = 0$ populates excited states
of $H_{g_1}$ according to the quench matrix (\ref{QuenchMatrix}).  
Letting the system evolve up to time $t_q$ and then `dequenching' back to $g_0$
results in a nontrivial amplitude of occupation for eigenstates of $H_{g_0}$, 
given by the quench propagator
\begin{eqnarray}
P_{\beta \a} (t_q) = 
\sum_{\gamma \in {\cal H}_{g_1}} Q_{g_0 g_1}^{\b \gamma} Q_{g_1g_0}^{\gamma \a}
e^{-i \omega_{g_1}^{\gamma} t_q},
\label{eq:doublequench}
\end{eqnarray}
where $\a, \b \in {\cal H}_{g_0}$ are respectively the labels for the pre- 
and post-quench states.  For $t_q = 0$, this propagator falls back onto the
identity matrix.  For finite duration $t_q > 0$, interference effects lead
to nontrivial  states (see left panel of Fig. \ref{DQidea}).
For a specific initial state $\a$ and final state
$\beta$, the quench propagator can be visualized as the sum of arrows of
length $|Q_{g_0 g_1}^{\beta \gamma} Q_{g_1 g_0}^{\gamma \a}|$ rotating as a 
function of $t_q$ at frequency $\omega_{g_1}^{\gamma}$ from an initial phase 
$\mbox{arg} (Q_{g_0 g_1}^{\beta \gamma} Q_{g_1 g_0}^{\gamma \a})$.
When arrows of non-negligible length align, constructive interference occurs, 
favoring the weight of the final state $\beta$ (see right panel of
Fig. \ref{DQidea}). Since all arrows rotate 
at different frequency, the occupation probability of state $\beta$ is a 
highly nontrivial function of the quench time, which is however completely 
characterized from the information we now have at hand. 

\begin{figure}[t]
\includegraphics[width=8cm]{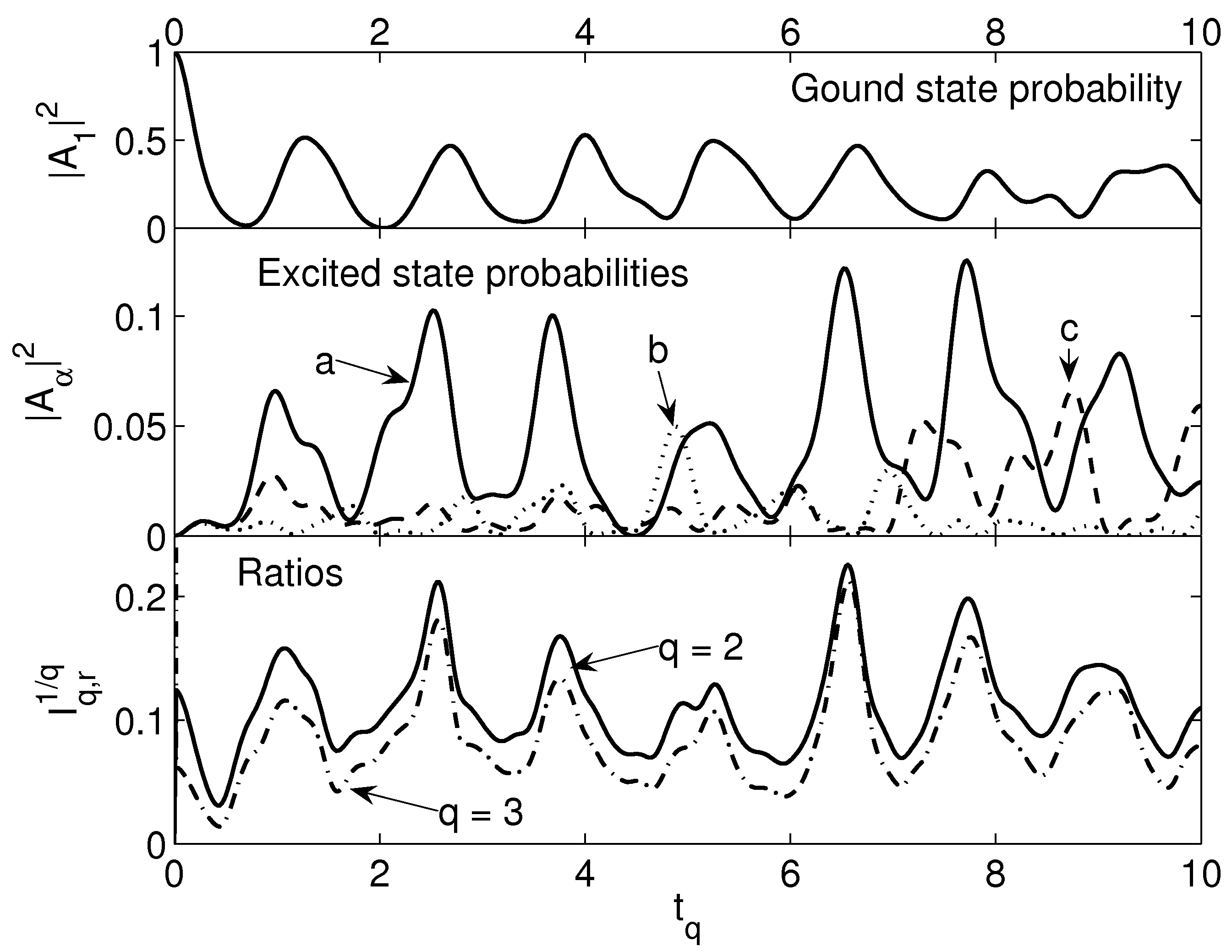}
\caption{Occupation probabilities and moments after a double quench between 
$g = 0$ and $0.5$, as a function of the quench duration $t_{q}$, for a system 
of 16 spins.  The top plot gives the ground state occupation probability,
%starting from 1 at $t_{q} = 0$, 
clearly displaying vanishing and reconstruction effects.  
The middle plot gives the occupation probabilities for the three more relevant 
states.
%described in the main text.  
The lower graph shows the moments $I_q$, quantifying the degree of
localization in Hilbert space after the double quench.}
\label{fig:doublequench_Occup}
\end{figure}

We consider for definiteness a double quench starting from the ground state of 
Hamiltonian $H_{g_0}$.
As a function of the quench duration $t_q$, the amplitudes of eigenstates $\a$ 
%\in {\cal H}_{g_0}$
after the dequench will thus be given by $A_{\a} (t_q) = P_{\a 0} (t_q)$.
We present in Fig. \ref{fig:doublequench_Occup} the results of such double 
quench calculations.
We specifically use a system of 16 spins, and trace over all intermediate 
states, allowing us to verify that the sum of square amplitudes remains equal
to one (up to numerical accuracy of order $10^{-7}$) at all quench times.  
The top panel shows the ground state occupation, which is inevitably the 
dominant state for small $t_q$.
However, we surprisingly find that its weight essentially vanishes (square 
amplitude below $0.005$), first around quench time $t_q = 0.56$, and also
repeatedly afterwards.  The ground state is also periodically reconstructed to
a large degree, showing that substantial sloshing of the occupation weight in 
the Hilbert space occurs as a function of the quench duration.

The occupation of individual excited states after the double quench also
displays prominent time-dependent interference effects.  
Their amplitudes all begin at zero for $t_q = 0$, but individual
states can attain non-negligible amplitudes when the `arrows' in their quench
propagator add up constructively for particular quench durations.  
This is shown in the middle panel of Fig. \ref{fig:doublequench_Occup}, 
where we plot the occupation probability of three example states among the 
single block states.  
The times at which such alignments take place can be predicted using a simple 
algorithm based on what could be called a continuous sieve of
Eratosthenes.  Namely, for a given final state $\beta$, the double quench
weights $|Q_{g_0 g_1}^{\beta \gamma} Q_{g_1 g_0}^{\gamma \a}|$ for all
$\gamma$ are first ordered in decreasing value.  The dominant mode (relabeled
$0$) has a time-dependent phase $\phi^0(t_q) = \omega_{g_1}^0 t_q - \phi^0$ 
with $\phi^0\equiv{\rm arg}(Q_{g_0g_1}^{\b\gamma^0} Q_{g_1 g_0}^{\gamma^0\a})$,
with similar defined phases for the subdominant modes $i > 0$.
Choosing an arbitrary phase alignment tolerance $\delta \theta$, the 
requirement that 
$|\phi^i(t_q) - \phi^0(t_q)| < \delta \theta$ for a given `arrow' $i$ defines 
excluded time intervals on the quench timeline $t_q \in [0, \infty[$.  
Erasing all such intervals for all states up to a level $n$
leaves only the times at which all phases $\phi^0(t_q), ..., \phi^n(t_q)$ 
are aligned to the chosen tolerance, and for which a certain amount of
constructive interference occurs.  For example, in the middle panel of
Fig. \ref{fig:doublequench_Occup}, the $b$ state peaks around $t_q \simeq4.8$;
it can be checked that this is a level $8$ alignment (with tolerance chosen as 
$\delta \theta = \pi/8$).
Alignments of a given order $n$ and tolerance $\delta \theta$ occur more or 
less periodically.  Increasing the order or reducing the tolerance $\delta
\theta$ makes alignments of higher quality but quickly increasing rarity. 
In view of this sieve of Eratosthenes logic, an interesting question is whether
the distribution of quench alignment times can be linked to that of {\it e.g.} 
prime numbers.

A study of the amplitudes $A_{\a}$ after a double quench for each individual
final state is clearly prohibitive.  Characterizing the distribution of
amplitudes is more enlightening, and is best performed by exploiting tools
common in the theory of localization in disordered systems, {\it i.e.} 
by considering the inverse participation ratios (IPRs)
$I_q = \sum_\a |A_{\a}|^{2q}$, with $I_1 = 1$.  
In the bottom panel of Fig. \ref{fig:doublequench_Occup}, we plot 
the second and third IPRs 
for excited states (defined as $I_{q,r}=\sum_{\a>0} |A_\a|^{2q}/
(\sum_{\a>0} |A_\a|^2)^q$, {\it i.e.} summing over excited states only), 
which display the localization tendencies of the excited states' amplitude 
weight in the Hilbert space after the double quench.  
Curves of $I_{2,r}$ and $I_{3,r}$ (always $\leq I_{2,r}$) approach one another 
when one excited state becomes dominant, and indicate smoother weight 
distribution otherwise.
%An interesting open problem is to investigate the full distribution of the
%IPRs, and investigate its (multi)fractality characteristics as a function of
%the quench strength and duration. 

\begin{figure}[t]
\includegraphics[width=8cm]{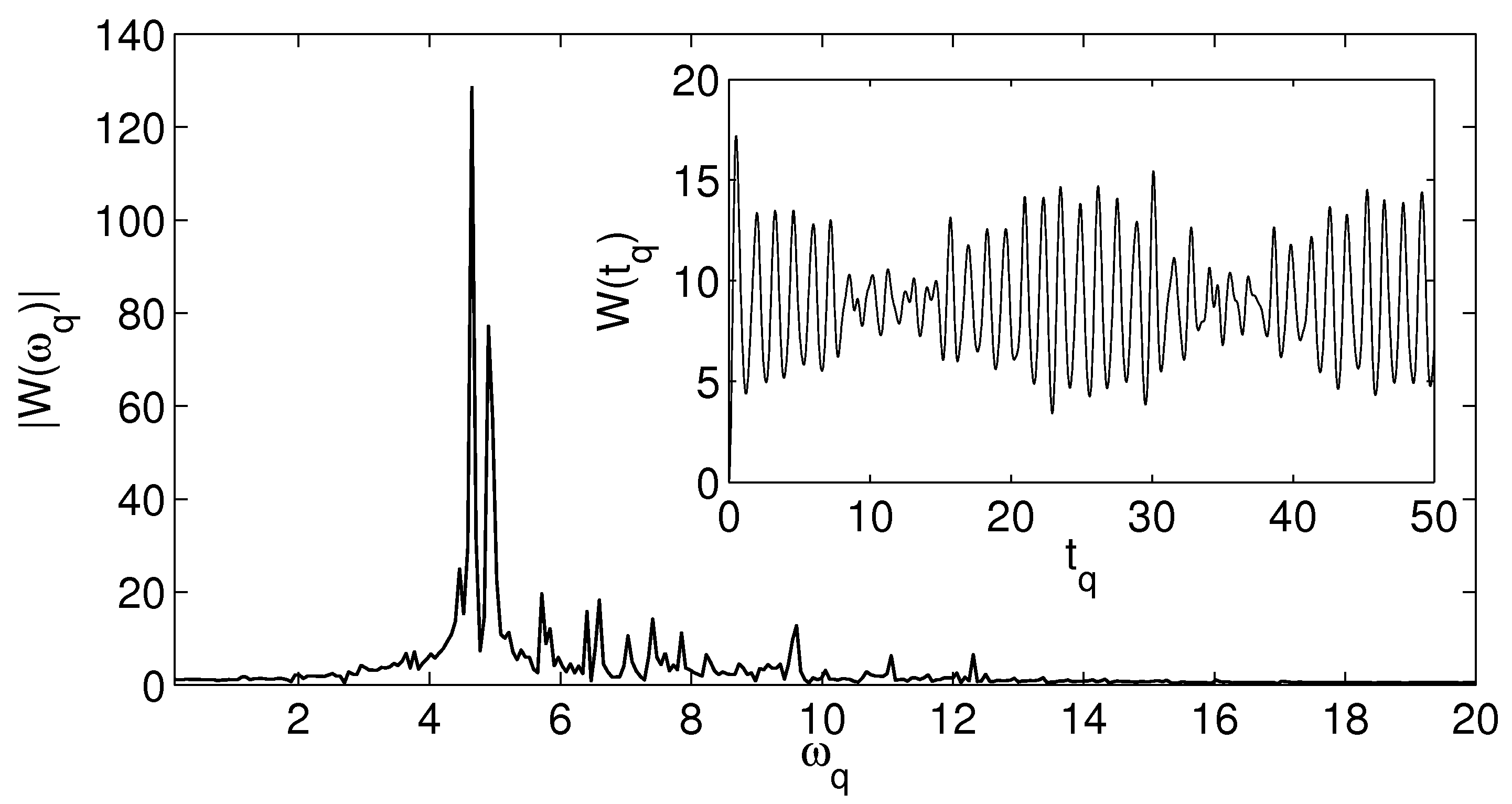}
\caption{Total energy (work) pumped into the system by the quench-dequench 
sequence with $g_0 = 0$ and $g_1 = 0.5$ and $16$ spins, as a function
of the quench duration $t_q$.  
Inset: work as a function of $t_q$ for the first oscillations of the envelope.
These persist for much longer times, which are not plotted for clarity.  
Main plot: Fourier transform of the work, showing the main peak associated 
to the energy difference of the two dominant intermediate states. 
%during the quench.
}
\label{fig:doublequench_Work}
\end{figure}

Another interesting quantity to look at, which also has the advantage of being 
more directly accessible in experiments, is the work
\be
W(t_q) = \sum_\a (\omega_{g_0}^\a - \omega_{g_0}^0) |A_\a(t_q)|^2
\ee
or in other words the energy which is pumped into the system by a
quench-dequench sequence of duration $t_q$.  Starting from the ground
state means that $W(t_q)$ is strictly positive. 
Since the quench-dequench sequence populates excited states in a highly $t_q$ 
dependent way, this quantity will also display a rich frequency profile.   
In Fig. \ref{fig:doublequench_Work}, we plot (inset) the work as a function 
of %the quench duration 
$t_q$, which displays nontrivial oscillatory behavior
dominated by a frequency $\omega \simeq 4.62$ corresponding to the energy
difference between the two dominant intermediate states during the quench.  
The Fourier transform $W(\omega_q)$
%= \int_0^{\infty} dt_q W(t_q) e^{-i\omega_q t_q}$
is plotted in the main part of the figure, clearly displaying the 
above-mentioned peak but also the non-negligible contributions from a broad
range of frequencies.  The position of the peaks corresponds to excited energy
level differences of the Hamiltonian during the quench, their height giving
information on the size of the relevant quench matrix elements.  The work
can thus be used not only as a spectroscopic tool, but as a way to quantify
eigenstate overlaps.

\section*{DISCUSSIONS}

In this work, we have proposed
a novel method to tackle quantum quenches, based on integrability. 
We applied the method to the fermionic pairing model, showing
the large amount of information that can be obtained on {\it e.g.} the
work probability density, physical observables (and correlation functions) 
and their time evolution, and multiple quenches settings.  
Everything has been derived in an exact (or numerically exact) manner in the 
mesoscopic regime, where quantum fluctuations govern the dynamics. 
We gave evidence of how single and double quench dynamics 
can be effectively used to extract spectroscopic data from simple 
measurable quantities like the work done on the system.

To obtain these results we explored the peculiar property that the quench 
matrix of the pairing model can be obtained using Slavnov's formula. 
This is not true in a general integrable model, and the generalization 
of the quench matrix representation is an open problem in the theory of 
integrable systems. 
When this representation will be available, the methods we propose here 
will allow exact calculations for a large variety of experimental relevant 
models, most importantly the one-dimensional Bose gas and Heisenberg 
spin chains.

{\it Acknowledgments.}
We thank Boris L. Altshuler and Jan von Delft for useful discussions. 
This work was supported by the DFG through SFB631, SFB-TR12 and the
Excellence Cluster "Nanosystems Initiative Munich (NIM)".
All the authors are thankful for support from the Stichting voor
Fundamenteel Onderzoek der Materie (FOM) in the Netherlands.
PC benefited of a travel grant from ESF (INSTANS activity).

\section*{METHODS}

\subsection*{Solving the Richardson equations}

Because of the blocking effect excluding singly occupied levels from the 
dynamics \cite{rs-62}, Richardson's model also has a pseudospin representation 
$S^-_\a= c_{\a\downarrow}c_{\a\uparrow}$,  
$S^+_\a=c^\dag_{\a\uparrow}c^\dag_{\a\downarrow}$, $S^z_\a= c^\dag_{\a\uparrow}c^\dag_{\a\downarrow}c_{\a\downarrow}c_{\a\uparrow} -1/2$. 
The Hamiltonian becomes
\be
H=\sum^N_{\a=1}
\e_\a S^z_\a -g \sum^N_{\a,\b=1} S^+_\a S^-_\b\,
\label{spinH}
\ee
where $N$ is the number of unblocked levels.

At $g=0$ 
%, for $N_r$ pair vacancies in $N$ energy levels $\e_\a$ with  only
%double degeneracy, 
the $\binom{N}{N_r}$ solutions
to the Richardson equations are trivial. They are given by
Eq. (\ref{RICHWF}) with the $N_r$ rapidities set to be strictly equal  
to one of the energies $\e_\a$. 
%Clearly, the GS in that limit is built by
%choosing the $N_r$ highest energy levels, i.e.
%$w_1 = \e_N\,, w_2 =\e_{N-1} \ \dots \ w_{N_r} = \e_{N-N_r+1}$.
Apart from a few particular cases with a small number of particles, the
Richardson equations are not solvable analytically when $g\neq 0$,
and one should solve them numerically. 
The solutions are such that every $w_j$ is
either a real quantity or forms, with another parameter
$w_{j'}$, a complex conjugate pair (CCP), i.e. $w_{j'}^* =w_j$.
The mechanism for the CCPs formation is very easy:
as interactions are turned on, all $w_j$ are real quantities for small  
enough
$g$, but at a certain critical value of the coupling $g^*_j$
two rapidities will be exactly equal to one of the energy levels  
($w_j=w_{j'} =
\e_{\gamma}(j)$) and for $g> g^*_j$, the two parameters that collapsed
will form a CCP at least for a finite interval in $g$.  The situation is
in fact rather intricate:  the values $g^*_j$ are implicit functions of
all other rapidities, and can only be read off a full solution of the  
Richardson
equations for a specific choice of state.  Moreover, CCPs can split  
back into real pairs, whose components can then re-pair with neighboring  
rapidities.
Different choices of the parameters $\e_\a$ and of their eventual
degenerations specify different models.
%In all the preceding sections everything was completely general
%(modulo having to take some extra precautions in the case of coinciding  
%levels $\e_\a$), but from now on we
We specialize to the case of equally spaced levels.
We make the choice to use $\e_\a =\a$ which sets the zero of energy and
implies that every energy will be given in units of the (pair)  
inter-level spacing.
Furthermore we consider only half-filling of the energy levels
$(N=2N_r)$, when the number of rapidities $N_r$ equals the number of
particles $N_p$, while in general in our notations $N_r+N_p=N$. 
At the precise value of $g$ at which a pair of rapidities $(w_j,w_{j'})$
collapse into a CCP $(w_j=w_{j'}=\e_\gamma(j))$, the Richardson  
equations (\ref{RICHEQ}) labelled $j$ and $j'$ will include two  
diverging terms whose sum remains finite. In order to be able to treat  
these points numerically, one can define the real variables,
$w_{1,j} \equiv  w_j + w_{j'}$ and 
$w_{2,j} \equiv (2\e_{\gamma}(j) - w_j - w_{j'})/(w_j-w_{j'})^2$.
As first discussed in \cite{r-66},
we need to know beforehand which rapidities will
form a CCP and at which $\e_\gamma(j)$ it will happen in order to use
this type of change of variables. Here we find the various solutions to the Richardson equations 
numerically by increasing $g$ by small steps starting from the solution 
at $g=0$ and can therefore predict at every step, the upcoming formation of CCPs.
As a consequence of this procedure, any given state at finite coupling can then be defined uniquely by the $g=0$ state  from which it emerges (and the actual value of $g$).
 
%Several different methods have been developed to work with the divergences
%appearing at the critical points. 
%We will use the method we proposed in Ref. \cite{fcc-08} to which we refer
%for more details. Since we find the various solutions to the Richardson equations 
%numerically by increasing $g$ by small steps starting from the solution 
%at $g=0$. 

\subsection*{Scalar products and Form factors}

In Bethe Ansatz solvable models, Slavnov's formula \cite{s-89} is an economical
representation of the overlap of an eigenstate $|\{w\}\rangle $ with a general 
Bethe state $|\{v\}\rangle$ constructed using the same operators, but for 
which the set of rapidities $\{ v \}$ does not fulfil the Bethe
equations. This overlap is given as a determinant of an $N_r$ by $N_r$ matrix, 
which in the problem at hand reads \cite{zlmg-02}:
\be
\langle \{w\}|\{v\}\rangle =
\frac {{\rm det}_{N_r} J(\{ v_a \},\{ w_b \})\,\prod^{N_r}_{a \neq b}(v_b-w_a)}
{\prod _{b <a} (w_b -w_a) \prod _{a <b} (v_b -v_a)}\,,
\label{slavnov}
\ee
with the matrix elements of $J$ given in Ref. \cite{zlmg-02,fcc-08},
{\small
\bea
J_{ab} &=& \frac {v_b -w_b}{v_a -w_b}
\left[ \sum^N_{\a=1} \frac {1}
{(v_a -\e _\a)(w_b -\e_\a)} \right. \nonumber\\ && \left.
-2\sum _{c\neq  a}^{N_r} \frac {1}{(v_a-v_c)(w_b -v_c)}
\right].
\eea}

The solution to the inverse problem allows a determinant 
representation for the necessary form factors \cite{zlmg-02}
{\small
\begin{eqnarray*}
\langle \{w\}|S^z_\a|\{v\}\rangle =
\prod^{N_r}_{a=1} \frac {(w_a - \e_\a)} {(v_a - \e_\a)} 
%\nonumber\\  \times
\frac { {\rm det}_{N_r} \left (\frac12 {\cal T}%(\{w\}, \{v\})
- {\cal Q} (\a) %, \{ w\}, \{ v\}) 
\right )}
{\displaystyle\prod _{b > a} (w_b -w_a) \prod _{b <a} (v_b -v_a)}\,,
\end{eqnarray*}
}
with the matrix elements of ${\cal T}, {\cal Q}$ given by
{\small
\bea
&&{\cal T}_{ab} =
\frac{2\displaystyle\prod ^{N_r}_{{c \neq a}} (w_c - v_b)}{w_a-v_b}
\left[\sum _{c \neq b} \frac1{(v_b-v_c)}-\sum _{c \neq a} \frac1{(v_b-w_c)} 
\right], 
\nonumber\\&&
{\cal Q}_{ab}(\a) = \frac {\prod _{c \neq b} (v_c-v_b)} {(w_a-\e_\a)^2}.
\eea}
We explicitly used the fact that both states are solutions to the Richardson 
equations in order to write the matrix elements of ${\cal T}$ in a more 
compact form than in previous publications \cite{zlmg-02,fcc-08}.

The form factors for $S^-_\a S^+_\b$ can be written as sum of $N_r$ 
determinants by generalizing the method of Ref. \cite{fcc-08} for $\{v\}=\{w\}$
starting from the double sums in Ref. \cite{zlmg-02}.
For $\a \neq \b$ we have 
\begin{eqnarray}
\langle \{v\}|S^-_\a S^+_\b|\{w\}\rangle &=&
\frac{\displaystyle\prod_{c}\left(\frac{v_c-\e_\a}{w_c-\e_\a}\right)\displaystyle\prod_{k\ne q}\left(w_k-w_q\right)}{
\displaystyle\prod_{b>a}(v_a-v_b)\prod_{a>b}(w_a-w_b)} \nonumber\\
&\times& \sum _{q=1}^{N_r} %\prod_{k\ne q}(w_k-w_q)\prod_{k\ne q}(v_k-v_q) 
\frac{w_q-\e_\a}{w_q-\e_\b}
{\rm det} {\cal J}^q_{\alpha,\beta}\,,
\end{eqnarray}
where, defining $A_{ab}=J_{a b}\displaystyle\prod_{c\ne b} (v_c-w_b)$, the matrix elements are given by
{\small
\begin{eqnarray}
{\cal J}_{ab} &=&A_{a b}- \frac{\prod_{k\ne b,q}(w_k-w_b)}
{\prod_{k\ne b+1,q}(w_k-w_{b+1})} \nonumber\\&&\times\frac{w_b-\e_\a}{w_{b+1}-\e_\a}
A_{a b+1},\ \ b<q-1,\nonumber\\
{\cal J}_{aq-1} &=&  A_{a q-1} +2\frac{(w_q-\e_\b)(w_{q-1}-\e_\a)}{
w_{q-1}-w_q}\prod_c\left(
\frac{v_c-\e_\b}{w_c-\e_\b} 
\right)
\nonumber\\&&\times
\prod_{k\ne q-1} (w_k-w_{q-1})
\frac{(2v_a-\e_\a-\e_\b)}{(v_a-\e_\a)^2(v_a-\e_\b)^2} 
,
\nonumber\\
{\cal J}_{aq} &=& 1/(v_a - \e_\a)^2, 
\nonumber\\
{\cal J}_{ab} &=&  A_{a b}, \ \  \ b > q\,.
\end{eqnarray}
}
For $\a=\b$ they are calculated using Hellmann-Feynman theorem as explained
in the main text.


\begin{thebibliography}{999}

\bibitem{exp} 
M. Greiner,   O.~Mandel, T.~W.~H\"ansch, \& I.~Bloch,
Collapse and revival of the matter wave field of a Bose-Einstein condensate,
Nature {\bf 419}, 51 (2002); \\
T. Kinoshita , T. Wenger, \& D. S. Weiss, 
A quantum Newton's cradle,
Nature {\bf 440}, 900 (2006); \\
L.E. Sadler, J. M. Higbie, S. R. Leslie, M. Vengalattore, 
\& D. M. Stamper-Kurn,
Spontaneous symmetry breaking in a quenched ferromagnetic spinor 
Bose-Einstein condensate,
Nature {\bf 443}, 312 (2006);\\ 
S. Hofferberth, I. Lesanovsky, B. Fischer, T. Schumm, \& J. Schmiedmayer,
Non-equilibrium coherence dynamics in one-dimensional Bose gases,
Nature {\bf 449}, 324 (2007);\\
C. N. Weiler, T. W. Neely, D. R. Scherer, A. S. Bradley, M. J. Davis \& 
B. P. Anderson, Spontaneous vortices in the formation of Bose-Einstein 
condensates, Nature {\bf 455}, 948 (2008).

\bibitem{th-mix} 
M. Cramer, C.M. Dawson, J. Eisert, \& T.J. Osborne,
Quenching, relaxation, and a central limit theorem for quantum lattice systems,
Phys. Rev. Lett. {\bf 100}, 030602 (2008);\\
C. Kollath, A. Laeuchli, \& E. Altman,
Quench dynamics and non equilibrium phase diagram of the Bose-Hubbard model,
Phys. Rev. Lett. {\bf 98}, 180601 (2007);\\
S. R. Manmana, S. Wessel, R. M. Noack, \& A. Muramatsu,
Strongly correlated fermions after a quantum quench,
Phys. Rev. Lett. 98, 210405 (2007);\\
M. Rigol, V. Dunjko, V. Yurovsky, \&  M. Olshanii,
Relaxation in a completely integrable many-body quantum system: An ab initio
study of the dynamics of the highly excited states of lattice hard-core bosons,
Phys. Rev. Lett. {\bf 98}, 050405 (2007);\\
T. Barthel T \& U. Schollwock, 
Dephasing and the steady state in quantum many-particle systems
Phys. Rev. Lett. {\bf 100}, 100601 (2008);\\
M. Cramer, A. Flesch, I. P. McCulloch, U. Schollwoeck, \& J. Eisert,
Exploring local quantum many-body relaxation by atoms in optical superlattices,
Phys. Rev. Lett. {\bf 101}, 063001 (2008);\\
M. Rigol, V. Dunjko, \& M. Olshanii,
Thermalization and its mechanism for generic isolated quantum systems,
Nature {\bf 452}, 854 (2008);\\
A. Laeuchli \& C. Kollath, 
Spreading of correlations and entanglement after a quench in the
one-dimensional Bose-Hubbard model, 
J. Stat. Mech. (2008) P05018;  \\
P. Barmettler, A. M. Rey, E. Demler, M. D. Lukin, I. Bloch, \& V. Gritsev,
Quantum many-body dynamics of coupled double-well superlattices,
Phys. Rev. A {\bf 78}, 012330 (2008);\\ 
A. Flesch, M. Cramer, I.P. McCulloch, U. Schollwoeck, \& J. Eisert, 
Probing local relaxation of cold atoms in optical superlattices,
Phys. Rev. A {\bf 78}, 033608 (2008);\\
G. Roux, On quenches in non-integrable quantum many-body systems: the
one-dimensional Bose-Hubbard model revisited, 0810.3720;\\
P. Barmettler, M. Punk, V. Gritsev, E. Demler \& E. Altman, 
Relaxation of antiferromagnetic order in spin-1/2 chains following a quantum
quench, 0810.4845;\\
S. R. Manmana, S. Wessel, R. M. Noack, \& A. Muramatsu,
Time evolution of correlations in strongly interacting fermions after a
quantum quench, 0812.0561.



\bibitem{cc-06}
P. Calabrese \& J. Cardy, 
Time-dependence of correlation functions following a quantum quench,
Phys. Rev. Lett. {\bf 96}, 136801 (2006);\\
P. Calabrese \& J. Cardy, 
Quantum quenches in extended systems, J. Stat. Mech. P06008 (2007).

\bibitem{gdlp-07} 
V. Gritsev, E. Demler, M. Lukin, \& A. Polkovnikov,
Spectroscopy of collective excitations in interacting low-dimensional
many-body systems using quench dynamics,
Phys. Rev. Lett. {\bf 99}, 200404 (2007).

\bibitem{IS}
E. Barouch, B. McCoy, \& M. Dresden, 
Statistical mechanics of the XY model. I,
Phys. Rev. A {\bf 2}, 1075 (1970); \\
E. Barouch \& B. McCoy, 
Statistical mechanics of the XY model. III, 
Phys. Rev. A {\bf 3}, 3127 (1971);\\
F. Igloi \& H. Rieger, 
Long-Range correlations in the nonequilibrium quantum relaxation 
of a spin chain,
Phys. Rev. Lett. {\bf 85}, 3233 (2000);\\
K. Sengupta, S. Powell, \& S. Sachdev, 
Quench dynamics across quantum critical points,
Phys. Rev. A {\bf 69}, 053616 (2004); \\
P. Calabrese \& J. Cardy, 
Evolution of entanglement entropy in one dimensional systems,
J. Stat. Mech. P04010 (2005);\\
R. W. Cherng \& L. S. Levitov, 
Entropy and correlation functions of a driven quantum spin chain,
Phys. Rev. A  {\bf 73}, 043614 (2006);\\
V. Mukherjee, U. Divakaran, A. Dutta, \& D. Sen,
Quenching dynamics of a quantum XY spin-1/2 chain in a transverse field,
Phys. Rev. B {\bf 76}, 174303 (2007);\\
V. Eisler \& I. Peschel, Entanglement in a periodic quench,
Ann. Phys. (Berlin) {\bf 17}, 410 (2008);\\
M. Fagotti \& P. Calabrese, 
Evolution of entanglement entropy following a quantum quench: Analytic results
for the XY chain in a transverse magnetic field, 
Phys. Rev. A {\bf 78}, 010306(R) (2008);\\
V. Mukherjee, A. Dutta, \& D. Sen, 
Defect generation in a spin-1/2 transverse XY chain under repeated quenching
of the transverse field, 
Phys. Rev. B {\bf 77}, 214427 (2008);\\
D. Rossini, A. Silva, G. Mussardo \& G. Santoro, 
Effective thermal dynamics following a quantum quench in a spin chain,
0810.5508.

\bibitem{s08}
A. Silva, The statistics of the work done on a quantum critical system by
quenching a control parameter, 
Phys. Rev. Lett. {\bf 101}, 120603 (2008).

\bibitem{BetheZP71}
H. Bethe,  Zur theorie der metalle, Z. Phys. {\bf 71}, 205 (1931).

\bibitem{KorepinBOOK}
V. E. Korepin, N. M. Bogoliubov \& A. G. Izergin A~G, {\em Quantum Inverse
  Scattering Method and Correlation Functions\/}, 
Cambridge University Press (1993).

\bibitem{TakahashiBOOK}
M. Takahashi, {\em Thermodynamics of one-dimensional solvable models\/},
  Cambridge University Press (1999).

\bibitem{s-89}
N. A. Slavnov, On scalar products in the algebraic Bethe ansatz,
Teor. Mat. Fiz. {\bf 79}, 232 (1989).

\bibitem{KitanineNPB554} 
N. Kitanine, J. M. Maillet, \& V. Terras, 
Correlation functions of the XXZ Heisenberg spin-1/2 chain in a magnetic field,
Nucl. Phys. B {\bf 554}, 647 (1999).

\bibitem{cm-05}
J.-S. Caux \& J.-M. Maillet, 
Computation of dynamical correlation functions of Heisenberg chains in a field,
Phys. Rev. Lett. {\bf 95}, 077201 (2005);\\
J.-S. Caux, R. Hagemans \& J.-M. Maillet,
Computation of dynamical correlation functions of Heisenberg chains: 
the gapless anisotropic regime,
J. Stat. Mech. P09003 (2005).

\bibitem{cc-06b}
J.-S. Caux \& P. Calabrese,
Dynamical density-density correlations in the one-dimensional Bose gas,
Phys. Rev. A {\bf 74}, 031605 (2006);\\
J.-S. Caux, P. Calabrese \& N. A. Slavnov,
One-particle dynamical correlations in the one-dimensional Bose gas,
J. Stat. Mech. P01008 (2007).


\bibitem{rs-62}
R. W. Richardson, 
A restricted class of exact eigenstates of the pairing-force Hamiltonian,
Phys. Lett. {\bf 3}, 277 (1963);\\
R. W. Richardson, Application to the exact theory of the pairing model to some
even isotopes of lead, 
Phys. Lett. {\bf 5}, 82 (1963); \\
R. W. Richardson \& N. Sherman, 
Exact eigenstates of the pairing-force Hamiltonian,
Nucl. Phys. {\bf 52}, 221 (1964);\\
R. W. Richardson \& N. Sherman, 
Pairing models of Pb$^{206}$, Pb$^{204}$ and Pb$^{202}$,
Nucl. Phys. {\bf 52}, 253 (1964).


\bibitem{rev}
J. von Delft \& D. C. Ralph, Spectroscopy of discrete energy levels in
ultrasmall metallic grains,  
Phys. Rep. {\bf 345}, 61 (2001);\\
J. Dukelsky S. Pittel \& G. Sierra, 
Exactly solvable Richardson-Gaudin models for many-body quantum systems,
Rev. Mod. Phys. {\bf 76},  643 (2004).


\bibitem{bcs-57}
J. Bardeen, L. N. Cooper \& J. R. Schrieffer,
Microscopic theory of superconductivity,
Phys. Rev. {\bf 106}, 162 (1957);  \\
J. Bardeen, L. N. Cooper \& J. R. Schrieffer,
Theory of superconductivity, 
Phys. Rev. {\bf 108}, 1175 (1957).

\bibitem{zlmg-02}
H.-Q. Zhou J. Links, R.H. McKenzie, \& M.D. Gould,
Superconducting correlations in metallic nanoparticles: exact solution of the
BCS model by the algebraic Bethe ansatz, 
Phys. Rev. B {\bf 65}, 060502(R) (2002);\\
J. Links, H.-Q. Zhou, R.H. McKenzie, \& M.D. Gould, 
Algebraic Bethe ansatz method for the exact calculation of energy spectra and
form factors: applications to models of Bose-Einstein condensates and metallic
nanograins, 
J. Phys. A {\bf 36}, R63 (2003).

\bibitem{fcc-08}
A. Faribault, P. Calabrese, \& J.-S. Caux, 
Exact mesoscopic correlation functions of the pairing model,
Phys. Rev. B {\bf 77}, 064503 (2008).

\bibitem{bl-06}
R. A. Barankov \& L. S. Levitov,
Synchronization in the BCS pairing dynamics as a critical phenomenon,
Phys. Rev. Lett. {\bf 96}, 230403 (2006).

\bibitem{bcs}
E. A. Yuzbashyan, B. L. Altshuler, V. B. Kuznetsov, \& V. Z. Enolskii,
Nonequilibrium Cooper pairing in the nonadiabatic regime,
Phys. Rev. B {\bf 72}, 220503(R) (2005);\\
E. A. Yuzbashyan \& M. Dzero, 
Dynamical vanishing of the order parameter in a fermionic condensate,
Phys. Rev. Lett. {\bf 96}, 230404 (2006); \\
E. A. Yuzbashyan, O. Tsyplyatyev, \& B. L. Altshuler,
Relaxation and persistent oscillations of the order parameter in the
non-stationary BCS theory, 
Phys. Rev. Lett. {\bf 96}, 097005 (2006);\\
M. Dzero, E. A. Yuzbashyan, B. L. Altshuler, \& P. Coleman, 
Spectroscopic signatures of nonequilibrium pairing in atomic Fermi gases,
Phys. Rev. Lett. {\bf 99}, 160402 (2007); \\
R. A. Barankov \& L. S. Levitov, 
Excitation of the dissipationless Higgs mode in a fermionic condensate,
0704.1292;\\
A. Tomadin, M. Polini, M. P. Tosi, \& R. Fazio,
Nonequilibrium pairing instability in ultracold Fermi gases with population
imbalance, Phys. Rev. A {\bf 77}, 033605 (2008).
%E. A. Yuzbashyan \& O. Tsyplyatyev,
%The dynamics of developing Cooper pairing at finite temperatures,0712.4280;





\bibitem{largeg}
J. M. Roman, G. Sierra, \& J. Dukelsky,
Elementary excitations of the BCS model in the canonical ensemble, 
Phys. Rev. B {\bf 67}, 064510 (2003);\\
E. A. Yuzbashyan, A. A. Baytin, \& B. L. Altshuler,
Strong coupling expansion for the pairing Hamiltonian,
Phys. Rev. B {\bf 68}, 214509 (2003). 

\bibitem{ao-02}
A. Mastellone, G. Falci, \& R. Fazio,
A small superconducting grain in the canonical ensemble,
Phys. Rev. Lett. {\bf 80}, 4542  (1998);\\
L. Amico \& A. Osterloh, 
Exact correlation functions of the BCS model in the canonical ensemble,
Phys. Rev. Lett. {\bf 88}, 127003 (2002).

\bibitem{r-66}
R. W. Richardson, Numerical study of the 8-32-particle eigenstates of the
pairing Hamiltonian, Phys. Rev. {\bf 141}, 949 (1966).

\end{thebibliography}
\end{document}